\documentclass[5p,twocolumn]{elsarticle}
\pdfoutput=1

\usepackage{graphicx}

\usepackage{amssymb}

\begin{document}

\bibliographystyle{h-elsevier}


\begin{frontmatter}

\title{Scintillation Pulse Shape Discrimination in a Two-Phase Xenon Time Projection Chamber}

\author[Princeton]{J. Kwong\corref{cor1}}
\ead{jktwo@princeton.edu}
\author[Cleveland]{P. Brusov}
\author[Cleveland]{T. Shutt}
\author[Princeton]{C. E. Dahl}
\author[Cleveland]{A. I. Bolozdynya}
\author[Cleveland]{A. Bradley}

\address[Princeton]{Department of Physics, Princeton University, Princeton, NJ 08540,
USA.}
\address[Cleveland]{Department of Physics, Case Western Reserve University,
Cleveland, OH 44106, USA.}
\cortext[cor1]{Corresponding author}


\begin{abstract}
The energy and electric field dependence of pulse shape discrimination in liquid xenon have been measured in a 10~gm two-phase xenon time projection chamber.  We have demonstrated the use of the pulse shape and charge-to-light ratio simultaneously to obtain a leakage below that achievable by either discriminant alone.  A Monte Carlo is used to show that the dominant fluctuation in the pulse shape quantity is statistical in nature, and project the performance of these techniques in larger detectors.  Although the performance is generally weak at low energies relevant to elastic WIMP recoil searches, the pulse shape can be used in probing for higher energy inelastic WIMP recoils.	
\end{abstract}

\begin{keyword}
xenon two-phase detectors pulse shape discrimination

\PACS 95.30 Cq Elementary particle processes in Astrophysics,
95.35.+d Dark Matter, 25.40.Dn Elastic neutron scattering,
29.40.Mc Scintillator Detectors
\end{keyword}
\end{frontmatter}

\section{Introduction}
	
	Two-phase xenon time projection chambers have set competitive limits in the search for weakly interacting massive particles or WIMPs \cite{Angle:2007uj,Lebedenko:2008gb}.  These detectors have a liquid target and a gas region above with several photomultiplier tubes (PMTs) for detecting scintillation and electrodes biased at high voltages for establishing charge-collecting electric fields.  The typical event consists of two bursts of light: the primary scintillation (``S1'') and the proportional (electroluminescence) scintillation (``S2'') from the drifting electrons exciting the gas \cite{Bolozdynya:1999dp}.  Backgrounds can be removed in several ways.  With 3D position reconstruction capability, events at the edges of the detector and multiple scatter events can be removed, eliminating most events from external gammas and neutrons and surface contaminants.  The charge-to-light ratio of background electron recoils (from gammas and betas) is higher than that of nuclear recoils from WIMPs, allowing for the rejection of bulk betas and single Compton-scatters.  This method has a discrimination efficiency of $>$99\% down to 2~keVr (recoil energy) \cite{Angle:2007uj,Shutt:2007zz}.  The measurement of the S1 pulse shape also provides discrimination between recoil types, although generally at a lower efficiency than that afforded by the charge-to-light ratio.  Dark matter detectors using pulse shape discrimination (PSD) with NaI \cite{Smith:1996fu}, high pressure xenon gas \cite{Martoff:2007zz}, single-phase liquid xenon \cite{Alner:2005pa,Bernabei:1998ad}, neon \cite{Hime:2005ab}, and argon \cite{Boulay:2004deap}, and two phase-argon \cite{Benetti:2007cd,Rubbia:2005ge} have been employed or are currently under development, but thus far there has been no exploration of PSD in two-phase xenon time projection chambers.
	This paper explores the energy and electric field dependence of PSD in such a detector at energies relevant to dark matter searches.  Also explored is the use of the pulse shape quantity in conjunction with the charge-to-light ratio to achieve a background rejection efficiency beyond that achievable by either discriminant alone.  A pulse shape Monte Carlo is used to examine the fluctuations of the pulse shape discriminant and determine the efficiency of PSD in larger detectors, where the distribution in the time-of-flight of photons becomes important.

\section{Scintillation Mechanism}

	In condensed noble elements, both excitation and ionization produced by particle interactions eventually result in scintillation by the following processes:
	
1) De-excitation
\begin{eqnarray*}
	\textnormal{R}^* + \textnormal{R} &\rightarrow& \textnormal{R}^{*}_{2} \\
	\textnormal{R}^{*}_{2} &\rightarrow& 2 \textnormal{R} + h\nu
\end{eqnarray*}

2) Recombination
\begin{eqnarray*}
	\textnormal{R}^{+} + \textnormal{R} &\rightarrow& \textnormal{R}^{+}_{2} \\
	\textnormal{R}^{+}_{2} + e^{-} &\rightarrow& \textnormal{R}^{**} + \textnormal{R} \\
	\textnormal{R}^{**} &\rightarrow& \textnormal{R}^{*} + heat \\
	\textnormal{R}^* + \textnormal{R} &\rightarrow& \textnormal{R}^{*}_{2} \\
	\textnormal{R}^{*}_{2} &\rightarrow& 2 \textnormal{R} + h\nu
\end{eqnarray*}
The excimer (excited dimer) R$^{*}_{2}$ is presumed to exist in either the singlet or triplet state with measured decay times of 4 ns and 21 ns, respectively \cite{Hitachi:1983ab}.  While the decay times have been shown to be independent of the stopping power $dE/dx$, the ratio of singlet to triplet states $I_{1}/I_{3}$ has been observed to be positively correlated with $dE/dx$ \cite{Hitachi:1983ab,Kubota:1980ab}.  This provides a mechanism for PSD.  The $I_{1}/I_{3}$ of electron, alpha, and fission fragment interactions (ordered by increasing $dE/dx$) in liquid xenon have been measured to be 0.05, 0.45, and 1.6, respectively \cite{Hitachi:1983ab,Kubota:1978ab,Kubota:1982ab}.
For liquid xenon, a slow tail with a 40 ns fall time is present in electron recoils at zero electric field and disappears upon application of a field, which strongly suggests that the slow tail is due to charge recombination \cite{Hitachi:1983ab}.  The tail is absent in scintillation pulses of alpha and nuclear recoils, presumably because the recombination times in dense tracks are much lower than the excimer decay times \cite{Hitachi:1983ab,Kubota:1980ab}.  The conversion of singlets to triplets (the lower energy state) by free electrons has been proposed as the mechanism for the positive correlation between $dE/dx$ and $I_{1}/I_{3}$ \cite{Hitachi:1983ab} -- events with lower $dE/dx$ have electrons that recombine on longer time scales and consequently have a greater chance of converting states.  In this view, the longer recombination times associated with lower $dE/dx$ could be shaping the pulse in two ways -- through the singlet-to-triplet ratio and the recombination time.  Our measurement probes the pulse shape dependence on $dE/dx$ (by observing events of different energy) for a particular type of recoil as opposed to observing the correlation across different types of particle interactions as previous studies have done.

\section{Experimental Setup}

\begin{figure}[t]
	\centering 
	\includegraphics[width=8.5cm]{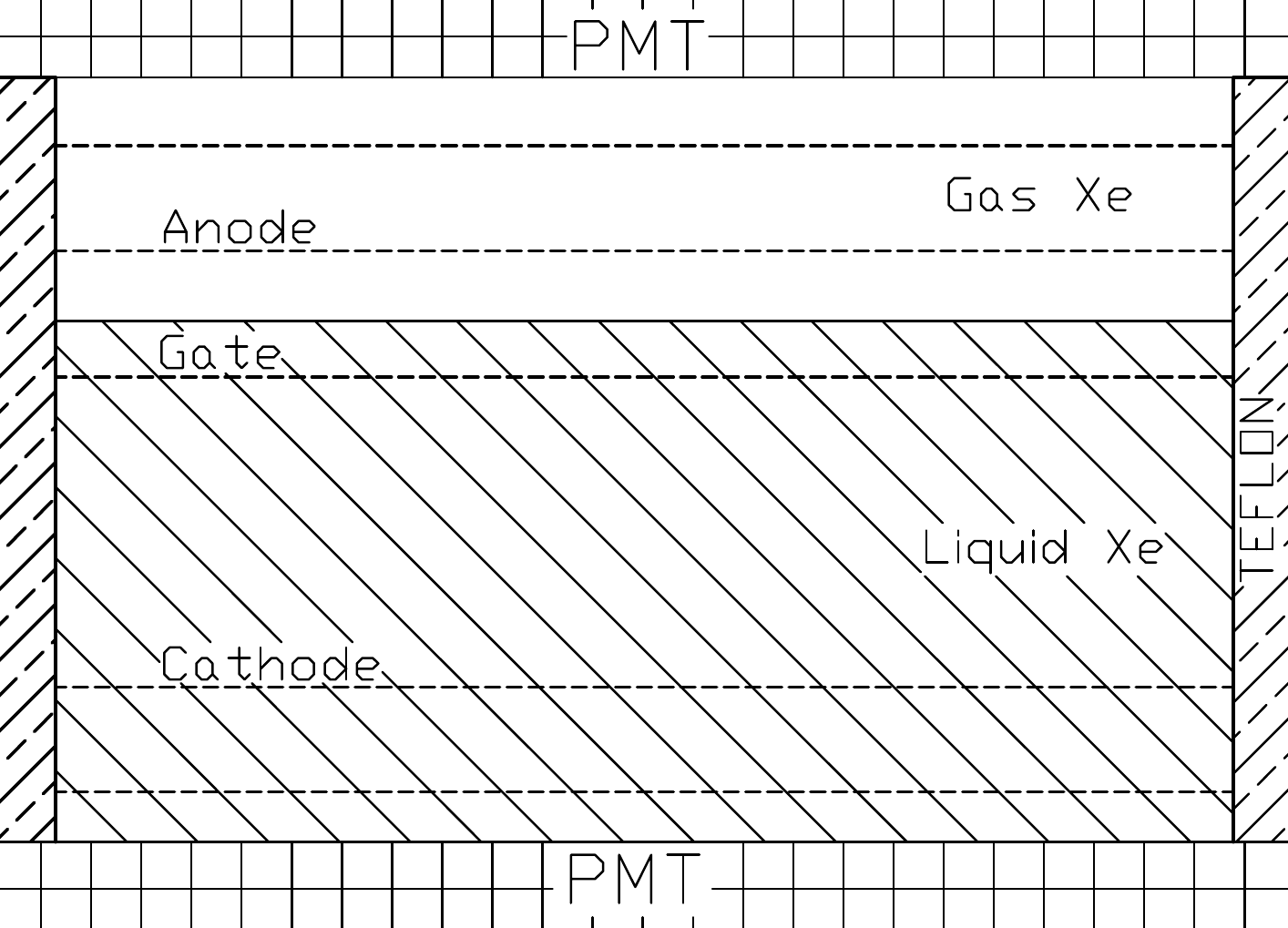}
	\caption{Diagram of the prototype with two PMTs and five wire grid electrodes.  The liquid surface sits midway between the gate and anode, which are separated by 4~mm.  The distance between the gate and cathode is 1~cm.}
	\label{fig:Xed1b_diagram}
\end{figure}

	The prototype detector (Figure \ref{fig:Xed1b_diagram}) consists of five wire grids and two Hamamatsu PMTs, one of which sits in the liquid (R9288) and the other in the gas (R6041).  The PMTs have quantum efficiencies of 16.8\% and 6.0\% at 175 nm, and collect 50\% and 10\% of the primary scintillation, respectively.  This results in an S1 signal size of $\sim$1 photoelectron (pe) per keVr for nuclear recoils and $\sim$5 pe per keV for electron recoils from 122 keV gammas at zero electric field.  (The light yield is roughly field independent for nuclear recoils and is very field dependent for electron recoils \cite{Aprile:2006kx}).  The PMT in the liquid collects more primary light than the other as most photons undergo internal reflection at the liquid surface.  The wire grids are parallel stretched wires soldered onto copper-patterned Cirlex boards.  The three lower grids in the liquid are made up of 40~micron beryllium copper wires.  The top two grids in the gas consist of 120~micron gold-plated aluminum wires.  All grids have a wire spacing of 2~mm.  The active volume contains 10~g of xenon and has a height of 1~cm and a diameter of 3.73~cm.  PTFE reflectors ($\geq$95\% reflectivity at $\sim$175 nm \cite{Yamashita:2004ptfe}) line the walls of the fiducial volume to boost the light signal and increase its spatial uniformity.

	By careful selection of materials, extensive cleaning of parts, baking and pumping of the detector, and recirculation of xenon through the detector and a hot metal getter purifier, we achieved an electron drift length of 30 cm (corresponding to a charge loss of $\leq$3\%) that remained constant throughout the experiment.  The detector is housed in a vacuum-insulated, liquid-nitrogen-cooled cold finger cryostat, which maintained a xenon temperature stable to within 0.1 K over several months.  Three parallel plate capacitors built onto the Cirlex rings monitor the liquid level, which was stable to 40 microns.

	The signals from the PMTs are passed through a 5$\times$ SRS SR445A amplifier and fan-in/out, and digitized at 500 MHz at a resolution of 8 bits.  Each PMT signal is digitized by two channels (one fine and one coarse) in order to extend the dynamic range from single photoelectrons to the saturation point of the amplifier.  The PMTs were measured to give a linear response in S1 and S2 over this range.  The DAQ triggers on S1 for events down to $\sim$5~keVr and on S2 for lower energy events with a threshold of $\sim$7 extracted electrons.  An energy threshold of 2 keVr is set in the analysis by a cut on S1.	

The data was taken with the liquid xenon temperature at 187.50~K with the liquid surface sitting midway between the anode and gate electrodes (Figure \ref{fig:Xed1b_diagram}).  The field above the liquid was kept at 10~kV/cm, which was necessary for full liquid-to-gas electron extraction \cite{Guschin:1982ab}.  At each of five different drift fields -- 0.06, 0.52, 0.88, 1.95, and 3.96~kV/cm -- three types of data sets were taken -- electron recoil, nuclear recoil, and calibration -- using the external sources $^{133}$Ba, $^{252}$Cf, and $^{57}$Co, respectively.  The high energy gammas from the $^{252}$Cf were shielded with 8'' of lead.  A data set was also taken to calibrate the signal in terms of photoelectrons.

\section{Pulse Shape Analysis}

\begin{figure}[t]
	\centering
	\includegraphics[width=8.0cm]{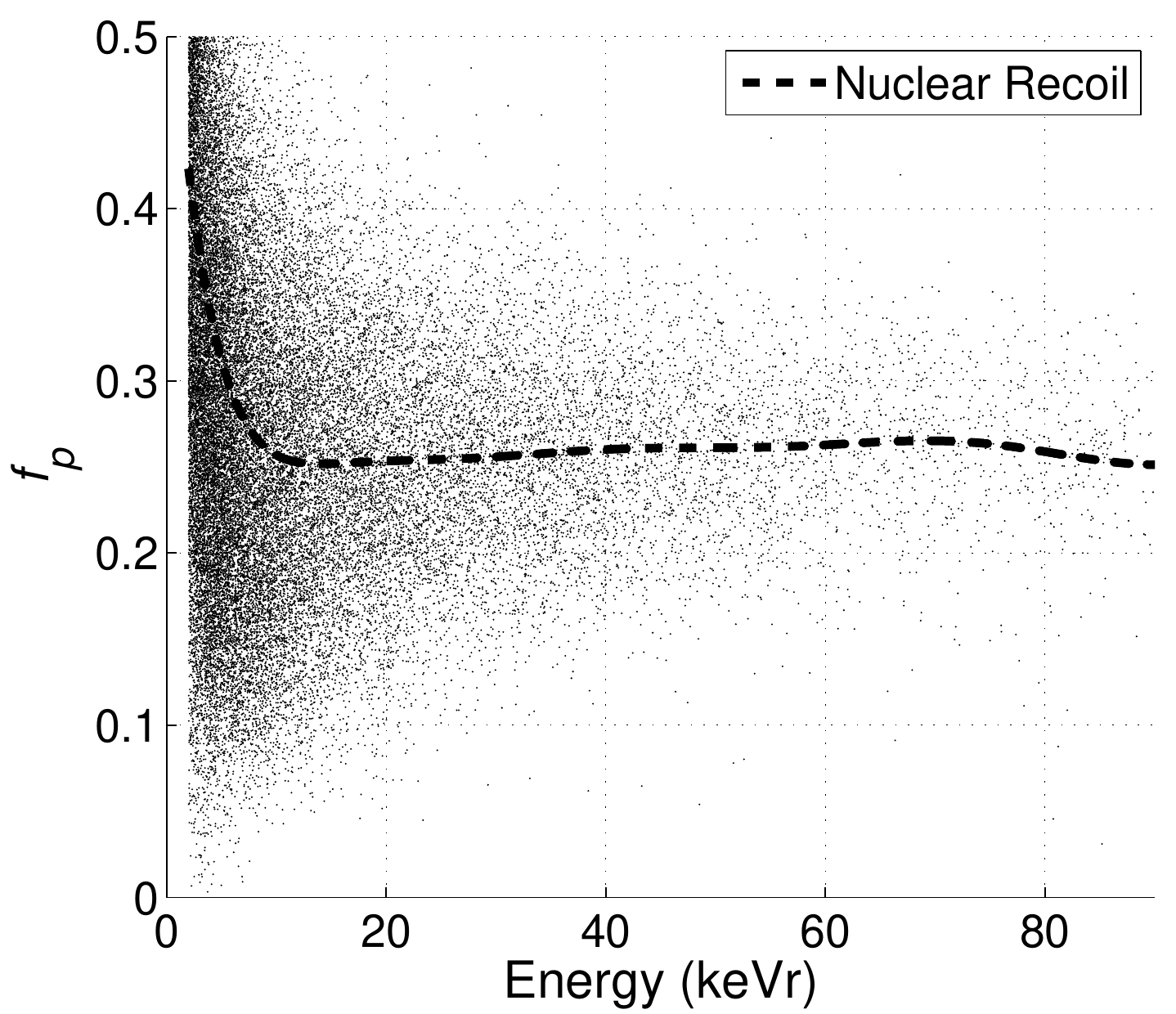}
	\includegraphics[width=8.0cm]{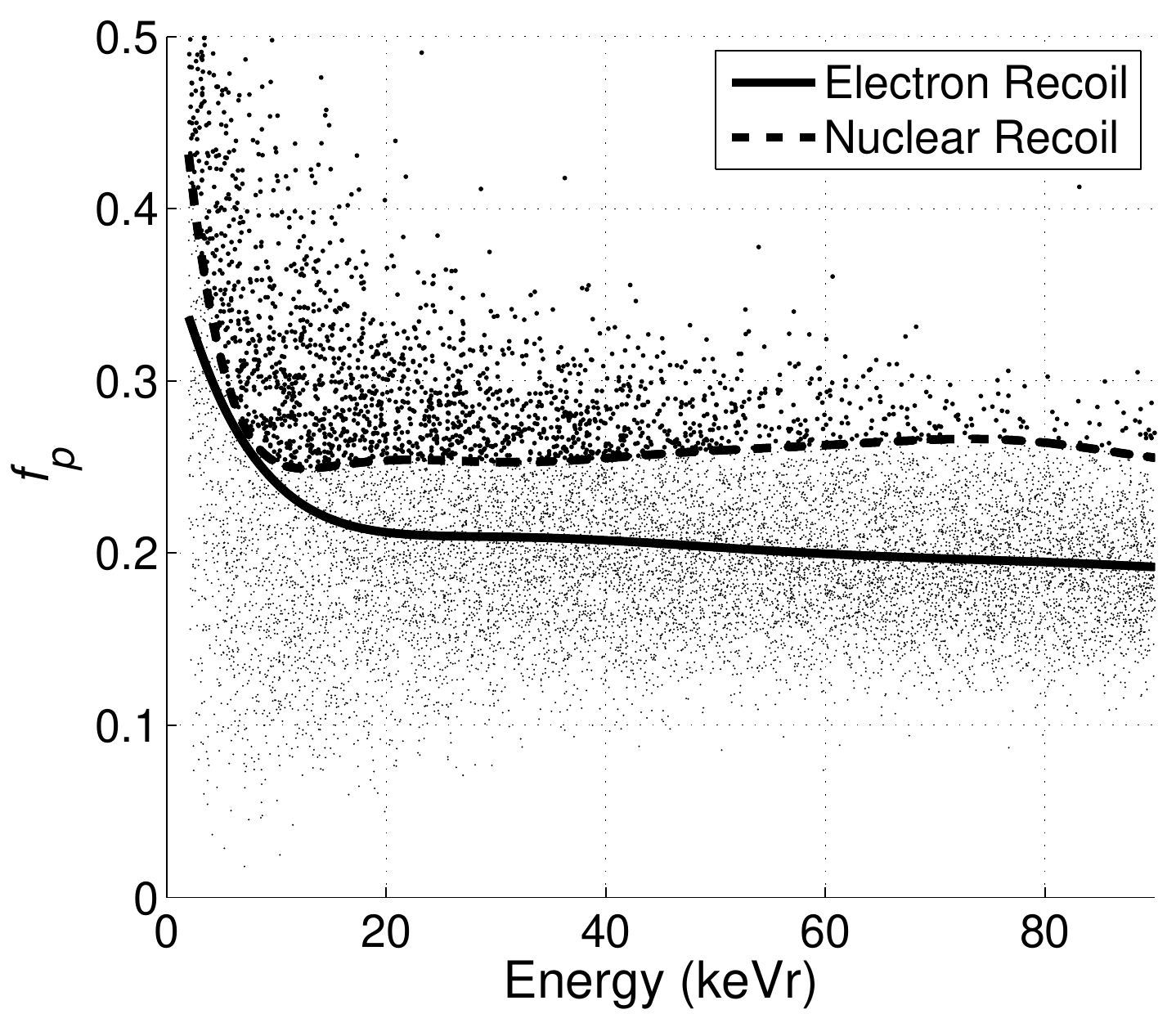}
	\caption{Scatter plots of $\it{f_p}$ versus recoil energy at 0.06 kV/cm.  Top~--~Nuclear recoils events from the $^{252}$Cf data set.  The electron recoil events from this data set have been removed with a cut in $\log_{10}(S2/S1)$.  Bottom~--~electron recoil events from the $^{133}$Ba data set with the leakage events highlighted.  The solid and dashed lines represent the electron and nuclear recoil means, respectively.}	
	\label{fig:ER_NR_prompt_total_versus_energy_0100_V_cm}
\end{figure}

	Data processing begins with the identification of scintillation pulses.	 Various pulse shape quantities such as the pulse height, integral, rise time, and width are calculated and used to classify each pulse as S1, S2 or noise.  Cuts are applied to remove all but single-scatter events (one S1 and one S2).  The integrals of the S1 and S2 pulses are corrected for depth dependence and calibrated with 122 keV gammas from $^{57}$Co.  The field dependence of the charge and light yields of these events have been measured previously \cite{Shutt:2007zz}.  The nuclear recoil energy is $E_r = E_e/\mathcal{L}_{eff} \cdot S_e / S_r$ where $E_e$ is the electron recoil energy scaled linearly from the S1 peak of 122~keV gammas; $S_e$ and $S_r$ are the electric field dependent scintillation yields of electron and nuclear recoils, respectively, relative to their values at zero field \cite{Aprile:2006kx}; and $\mathcal{L}_{eff}$ is the nuclear recoil scintillation yield relative to that of the 122 keV electron recoils at zero field.  The energy dependence of $\mathcal{L}_{eff}$ has been measured by several groups \cite{Bernabei:1998ad,Arneodo:2000icarusscint,Akimov:2001ukdmscint,Aprile:2005mt,Chepel:2006yv,Aprile:2008rc} and there is not yet a consensus on the correct value.  For simplicity, we assume a constant value of $\mathcal{L}_{eff} = 0.2$.
	
	A preliminary study of the PSD performance was done with various pulse shape quantities such as the width at half-max, integral/height, prompt/total (fraction of light at the head of the pulse), and a quantity called ``multi-bin'' which is a generalization of the prompt/total fraction to more than two bins (as described in \cite{Lippincott:2008ad}).  The last two quantities give equally the best performances.  We report results using prompt/total as it requires much simpler calculations.  The prompt/total is defined as
\begin{equation}
	\it{f_p} = \frac{\int^{t_0 + t_{window}}_{t_i} S(t) dt}{\int^{t_f}_{t_i} S(t) dt},
	\label{eqn:prompt_total}
\end{equation}
where $S(t)$ is the signal, $t_{window}$ is the prompt window size, $t_i = t_0 - 50$ ns, and $t_f = t_0 + 300$ ns, with $t_0$ as the time at which the pulse reaches 10\% of the maximum amplitude on the leading edge.  The optimal $t_{window}$ depends on the electronics and size of the detector, and is 10 ns for the prototype.
	Figure \ref{fig:ER_NR_prompt_total_versus_energy_0100_V_cm} shows a scatter plot of $\it{f_p}$ as a function of recoil energy for the nuclear recoils from the $^{252}$Cf data set.  The second scatter plot shows the electron recoils from $^{133}$Ba data set with the leakage events in the nuclear recoil acceptance window highlighted.  This window is defined as the region above the nuclear recoil $\it{f_p}$ mean, giving a nuclear recoil acceptance of 50\%.

	The Gaussian-extrapolated leakage is the fraction of electron recoil events in the nuclear recoil acceptance window as determined by the Gaussian width of the electron recoil $\it{f_p}$ and distance between the electron and nuclear recoil $\it{f_p}$ means.  For all fields, the leakage is calculated for discrete energy bins of edges 2, 6, 10, 20, 30, 40, 50, 60, 70, 80, and 90 keVr. The electron and nuclear recoil energy-dependent $\it{f_p}$ means are characterized by 10th-order polynomials (these are represented by the lines in Figures \ref{fig:ER_NR_prompt_total_versus_energy_0100_V_cm}).  These polynomials are calculated by iteratively fitting to the $\it{f_p}$ means of many overlapping energy bins, subtracting out the mean, and repeating until the mean-subtracted data has no more energy dependence.  Fitting to the flattened bands provides a more accurate measurement of the band width since a shift in the mean of $\it{f_p}$ within an energy bin would otherwise lead to an overestimate of the band width.  Before fitting to the nuclear recoils, a cut in charge-to-light ratio $\log_{10}(S2/S1)$ is applied to remove the electron recoil events.  The band separation is the separation according to the 10th-order polynomials plus the remaining offsets as determined by the Gaussian fits to the flattened bands.  The same leakage calculation procedure is used for $\log_{10}(S2/S1)$.  For this discriminant, the nuclear recoil acceptance window is below the $\log_{10}(S2/S1)$ mean.

\section{Pulse Shape Monte Carlo}

\begin{figure}[t]
	\centering
	\includegraphics[width=4.3cm]{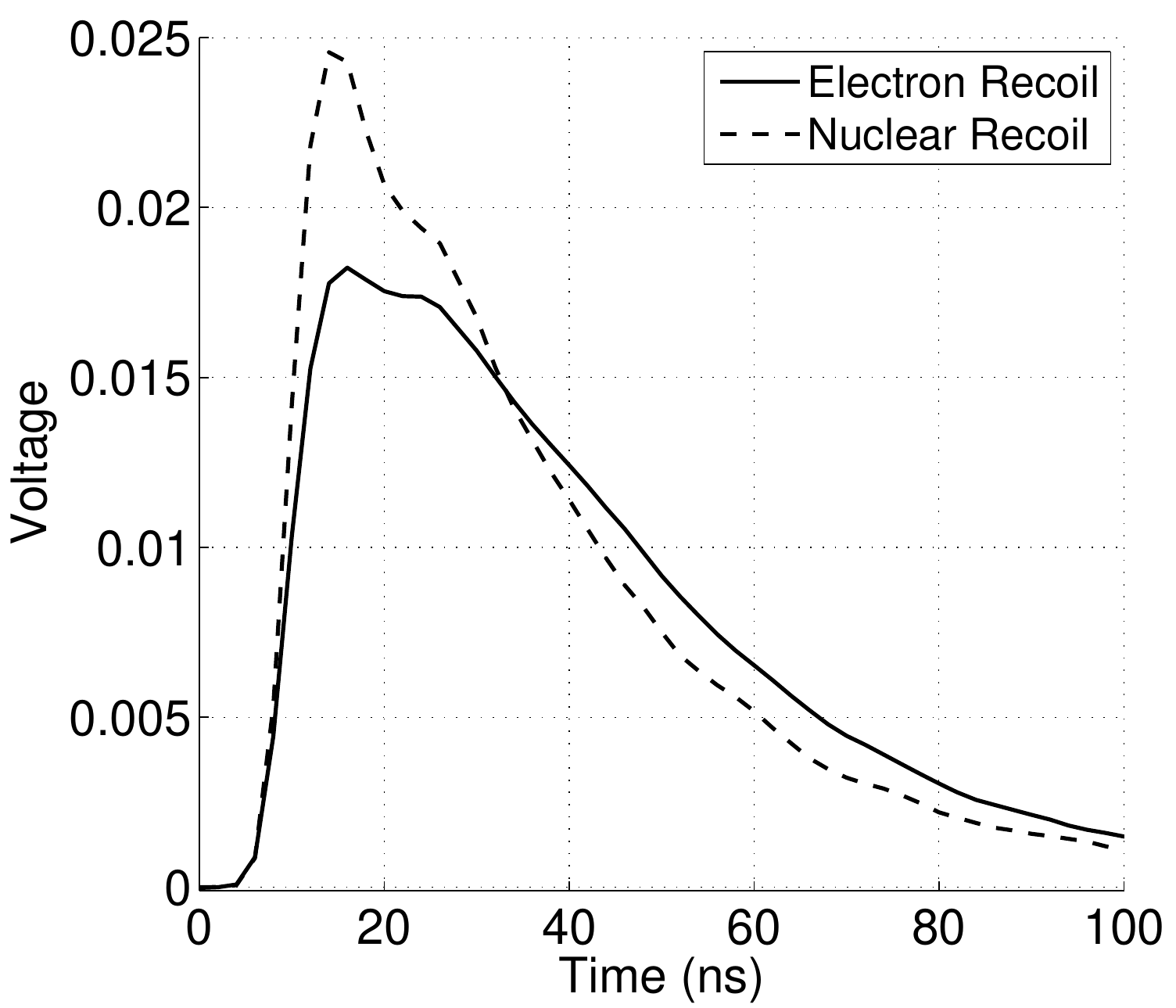}
	\includegraphics[width=4.3cm]{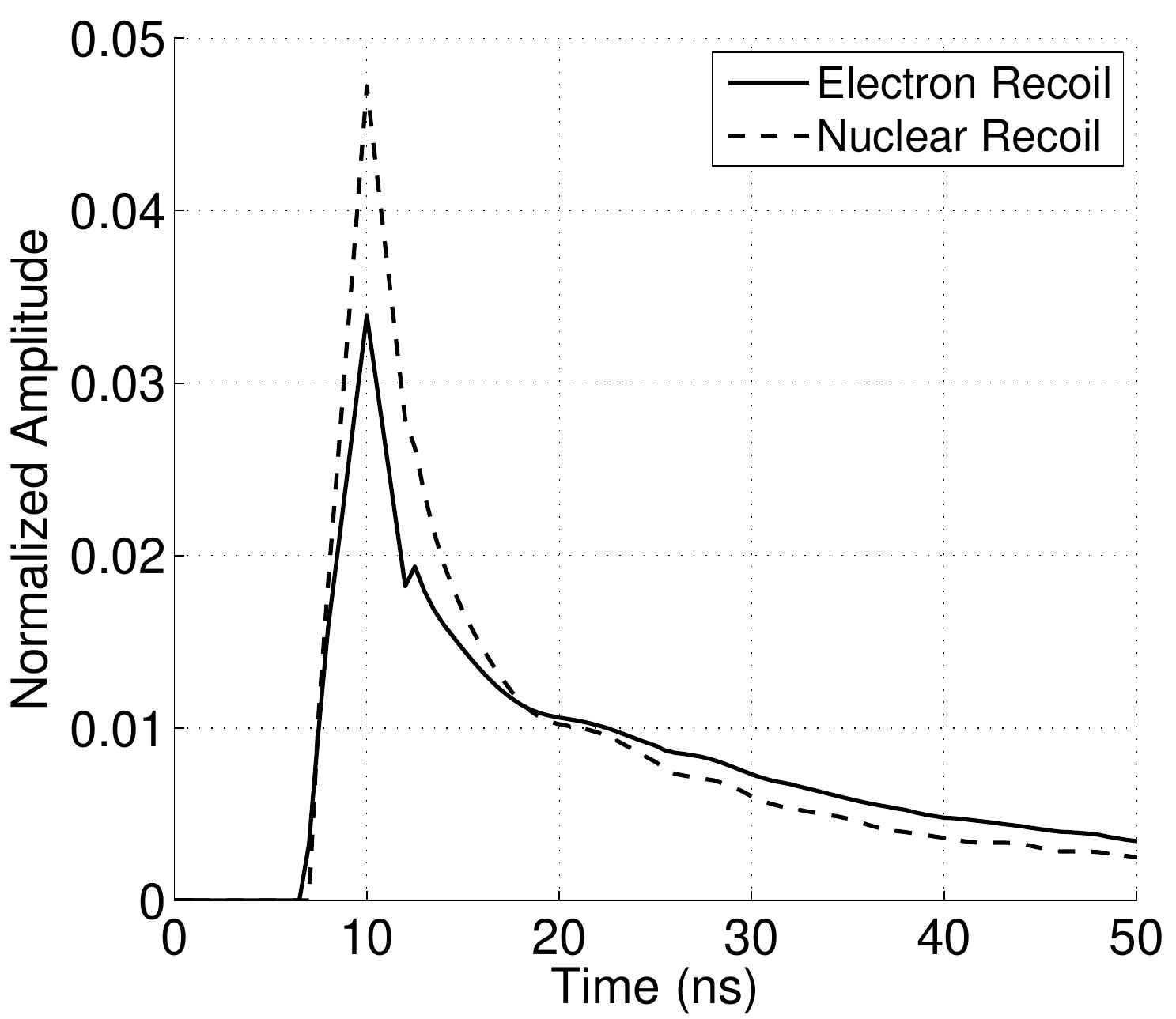}
	\caption{Average pulse (left) and probability distribution functions (right) for 78 keVr electron and nuclear recoils at 0.06 kV/cm.  The slight hump following the peak is an electronics glitch of unknown origin.}
	\label{fig:Decay_PDF_Filtered_AveragePulses_77_79_keVr_0100_V_cm}	
\end{figure}

\begin{figure}[tp]
	\centering
	\includegraphics[width=8.0cm]{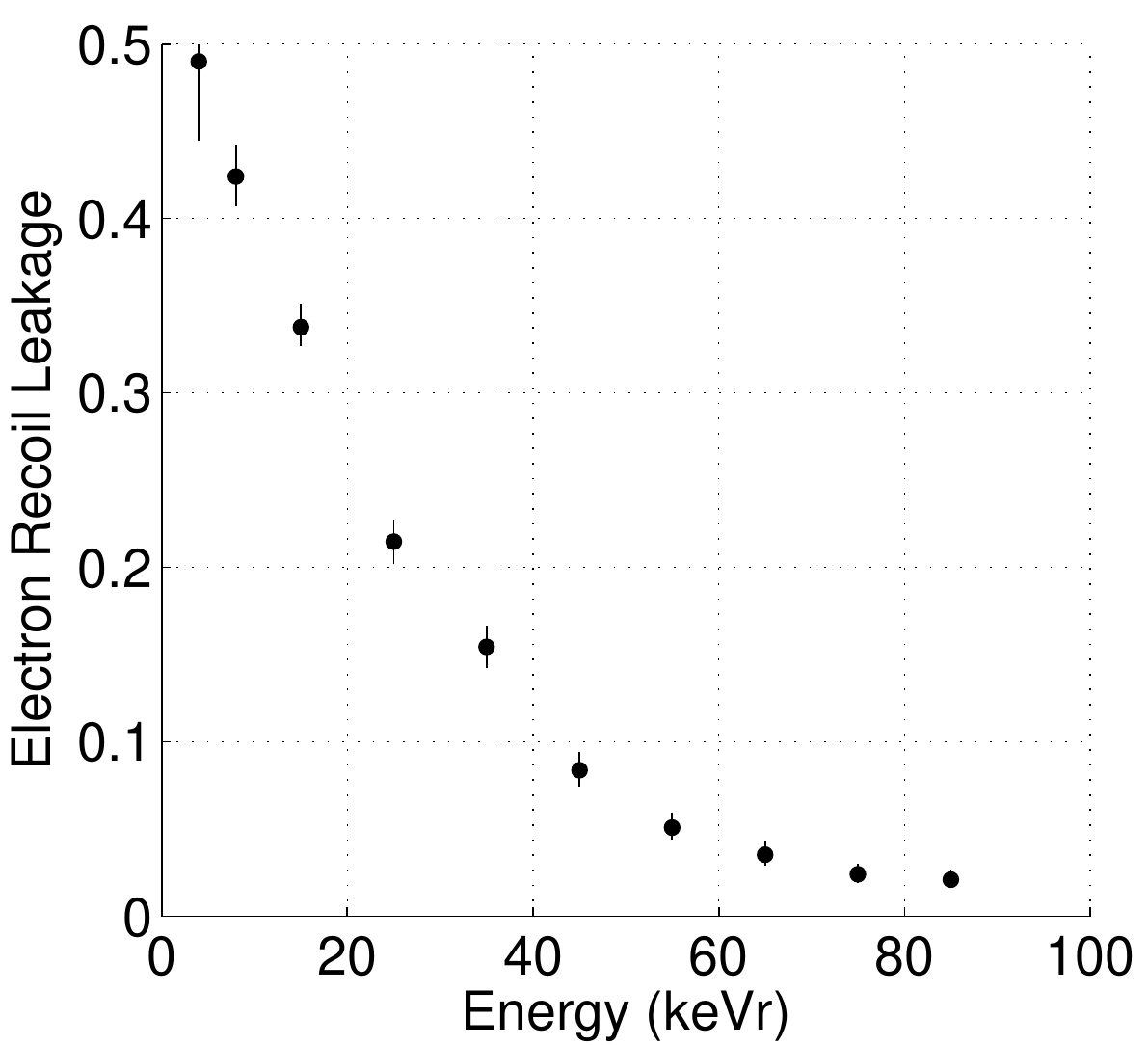}
	\caption{Gaussian-extrapolated electron recoil leakage as a function of recoil energy for the data set taken at a drift electric field of 0.06 kV/cm using the pulse shape discriminant $\it{f_p}$.}
	\label{fig:Xed_Leakage_Versus_keVr_0100_V_cm}
\end{figure}

\begin{figure}[t]
	\centering
	\includegraphics[width=8.0cm]{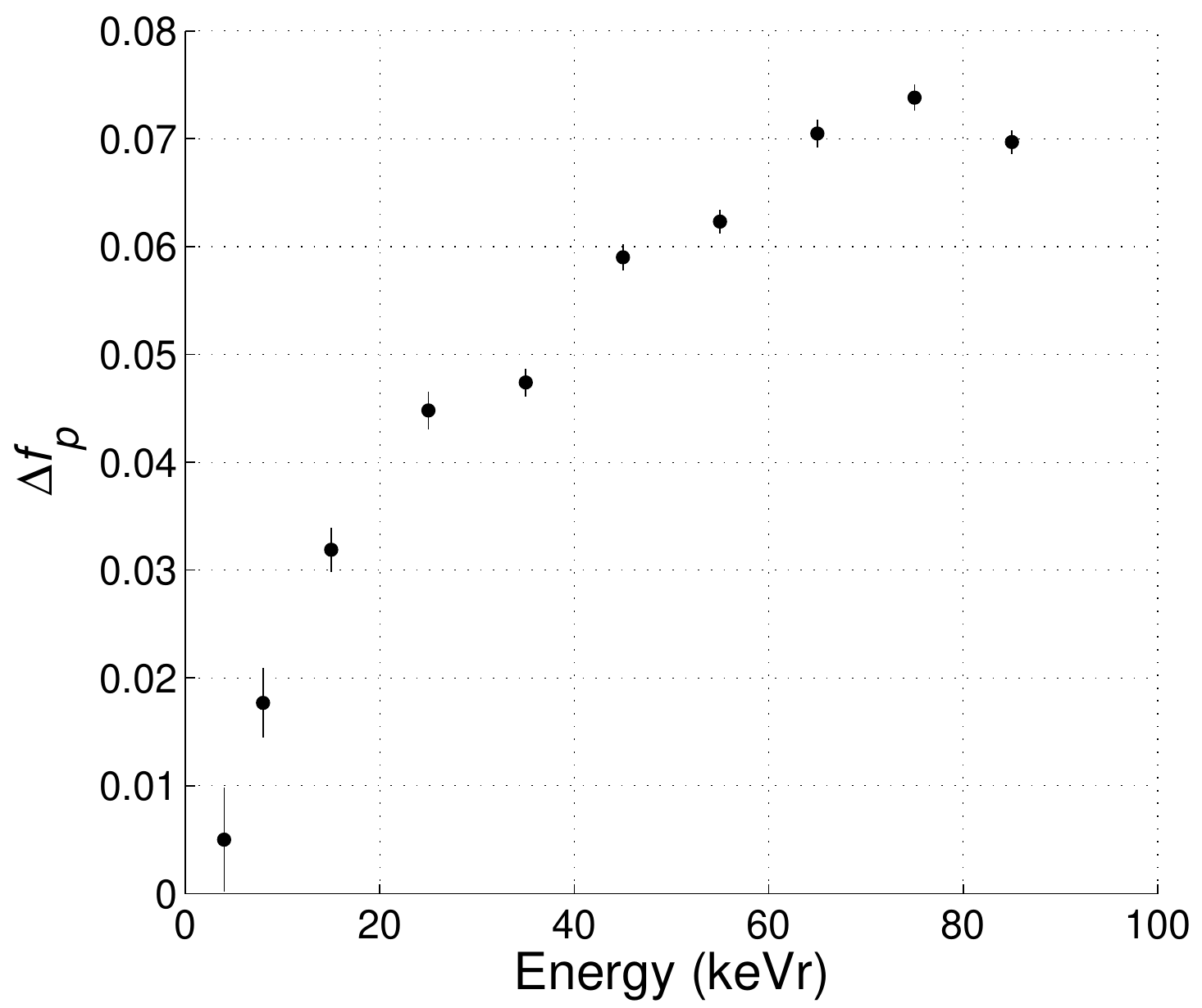}
	\includegraphics[width=8.0cm]{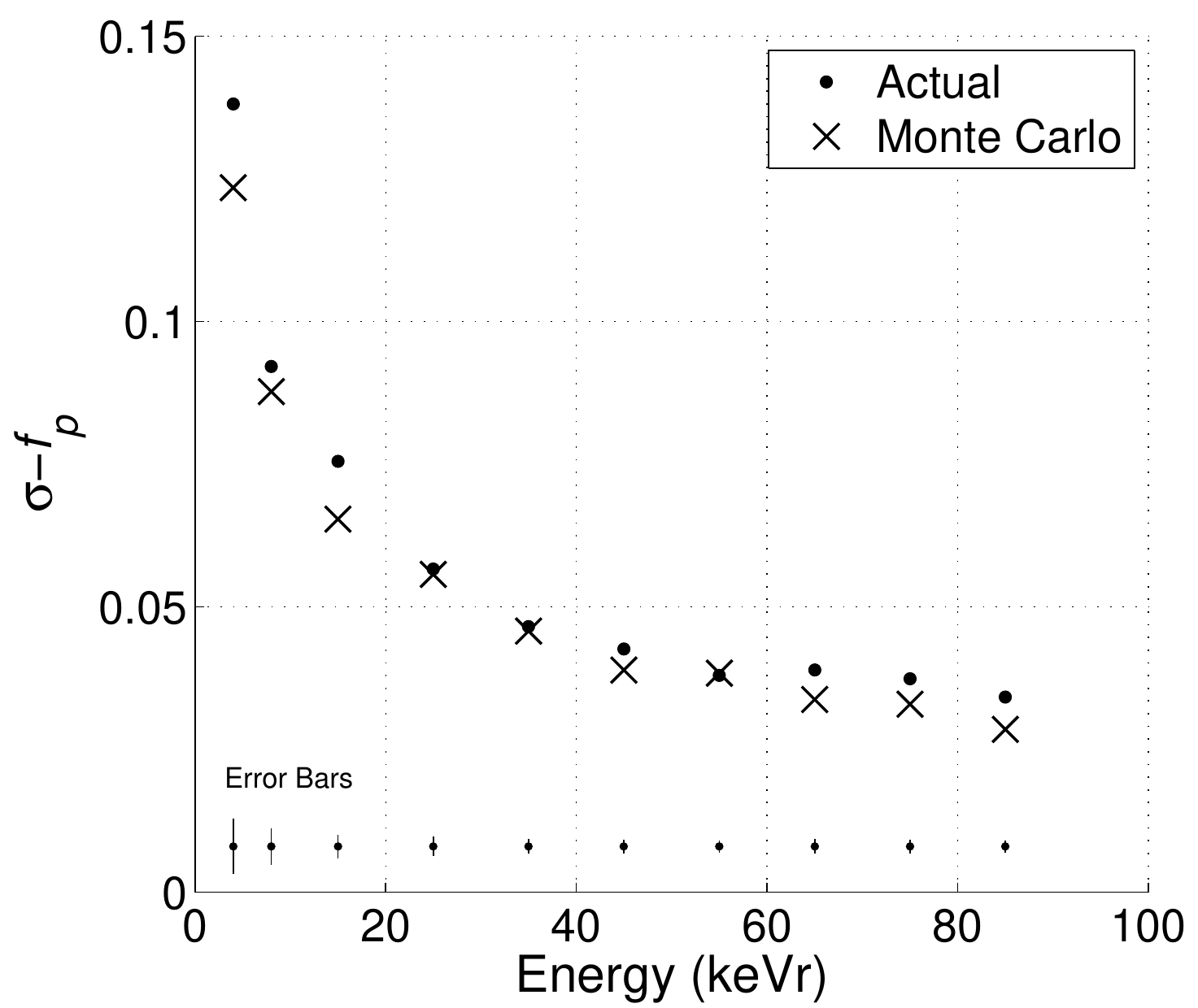}
	\caption{The energy dependence of the separation between electron and nuclear recoil $\it{f_p}$ fit means ($\Delta\it{f_p}$) (top) and electron recoil fit sigmas (bottom) at 0.06 kV/cm.  The error bars represent the statistical uncertainties.}
	\label{fig:band_separation_sigma_versus_energy_0100_V_cm}
\end{figure}

	A pulse shape Monte Carlo was built to gain a better understanding of the width of the $\it{f_p}$ distribution and to project the discrimination in larger detectors.  To simulate the S1 signal, several functions are needed: the photon emission (excimer decay) time probability distribution function (pdf), the photon time-of-flight pdf, PMT gain pdf, and average single photoelectron response waveform.  
	The photon time-of-flight pdf was obtained with a homemade photon propagation Monte Carlo.  The Rayleigh scattering and absorption lengths used in the Monte Carlo are 30 cm and 1 m for the liquid \cite{Baldini:2004ph,Ishida:1997ab}, and the corresponding values for gas are extrapolated by density.  The beryllium copper and gold-plated grids and PTFE walls are assumed to have reflectivities of 0.1, 0.2 and 0.95, respectively.  Although the metal reflectivities are not well known as they depend greatly on the surface condition, they have negligible effects on the final results.  The Monte Carlo shows that the time-of-flight is instantaneous for our purposes and thus is not included for simulating data for the prototype.  (However, time-of-flight pdfs are needed for the larger detectors discussed later.)  The PMT gain pdf is taken from the photoelectron calibration data set.  The photon emission pdfs are obtained by deconvolving the photoelectron response from the average S1 pulses.  The average pulses are calculated by lining up the pulses according to their $t_0$ and calculating the mean voltage for each time bin.  This was done for each 2 keVr bin from 2 to 90 keVr.  The average single photoelectron response waveform was obtained from the photoelectron calibration data set.  Figure \ref{fig:Decay_PDF_Filtered_AveragePulses_77_79_keVr_0100_V_cm} shows the average pulses and corresponding emission pdfs for 78 keVr electron and nuclear recoils from the data set taken at a drift field of 0.06 kV/cm.  Initially, the only objective of this experiment was to measure the discrimination efficiency of the charge-to-light ratio.  Consequently, little effort was made to match impedances for internal electrical lines, resulting in an ever-present glitch seen in the average pulses of Figure \ref{fig:Decay_PDF_Filtered_AveragePulses_77_79_keVr_0100_V_cm}.  This glitch gives rise to a sharp spike in the photon emission time pdf at about 25 ns and has been removed.

\begin{figure}[t]
\centering
\includegraphics[width=8.0cm]{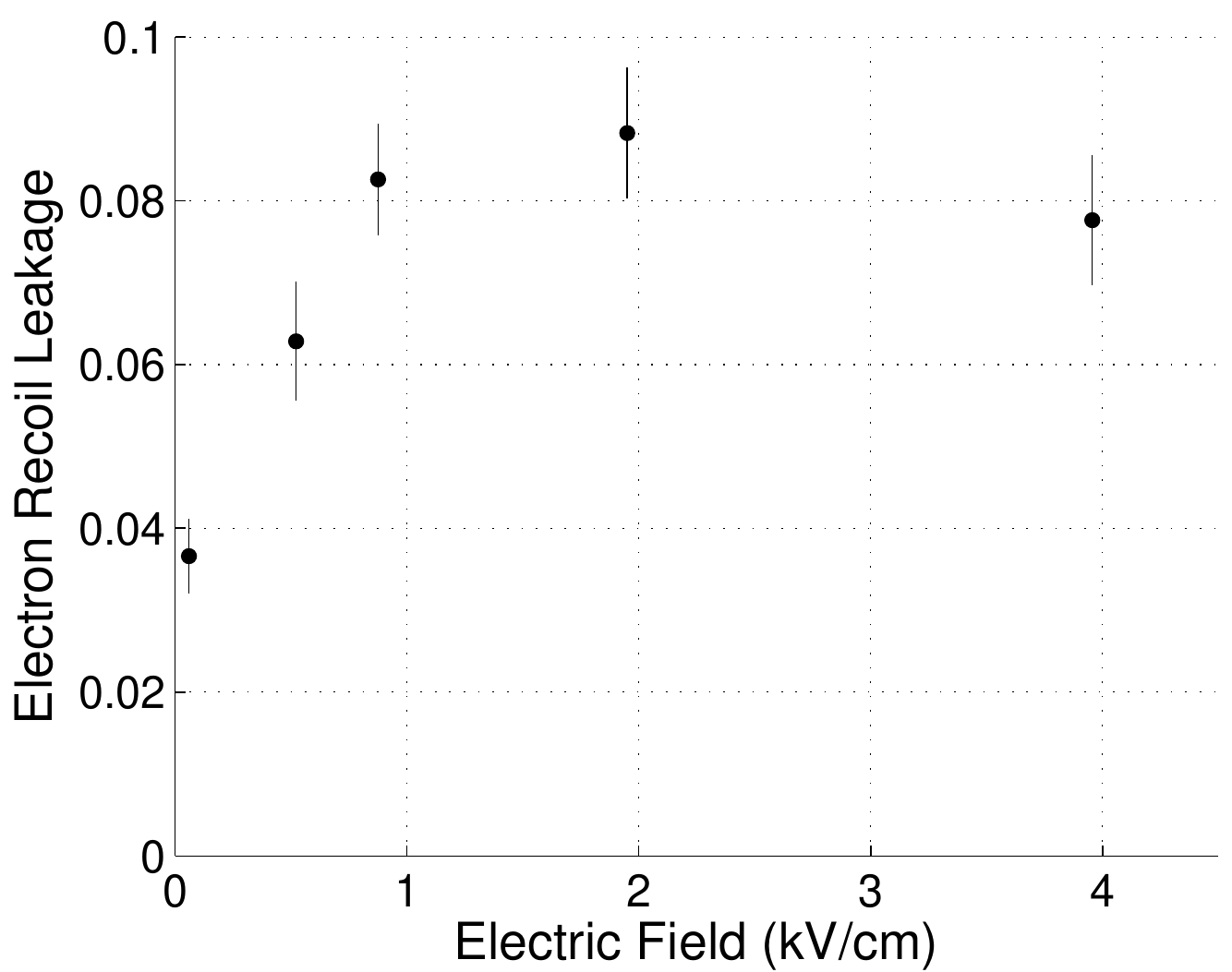}
\caption{Electron recoil leakage versus electric field between 60 and 80 keVr using $\it{f_p}$.  The error bars represent the statistical uncertainties.}
\label{fig:Leakage_Vs_Field_prompt_total_60_to_80_keVr}
\end{figure}

\begin{figure}[t]
	\centering
	\includegraphics[width=8.0cm]{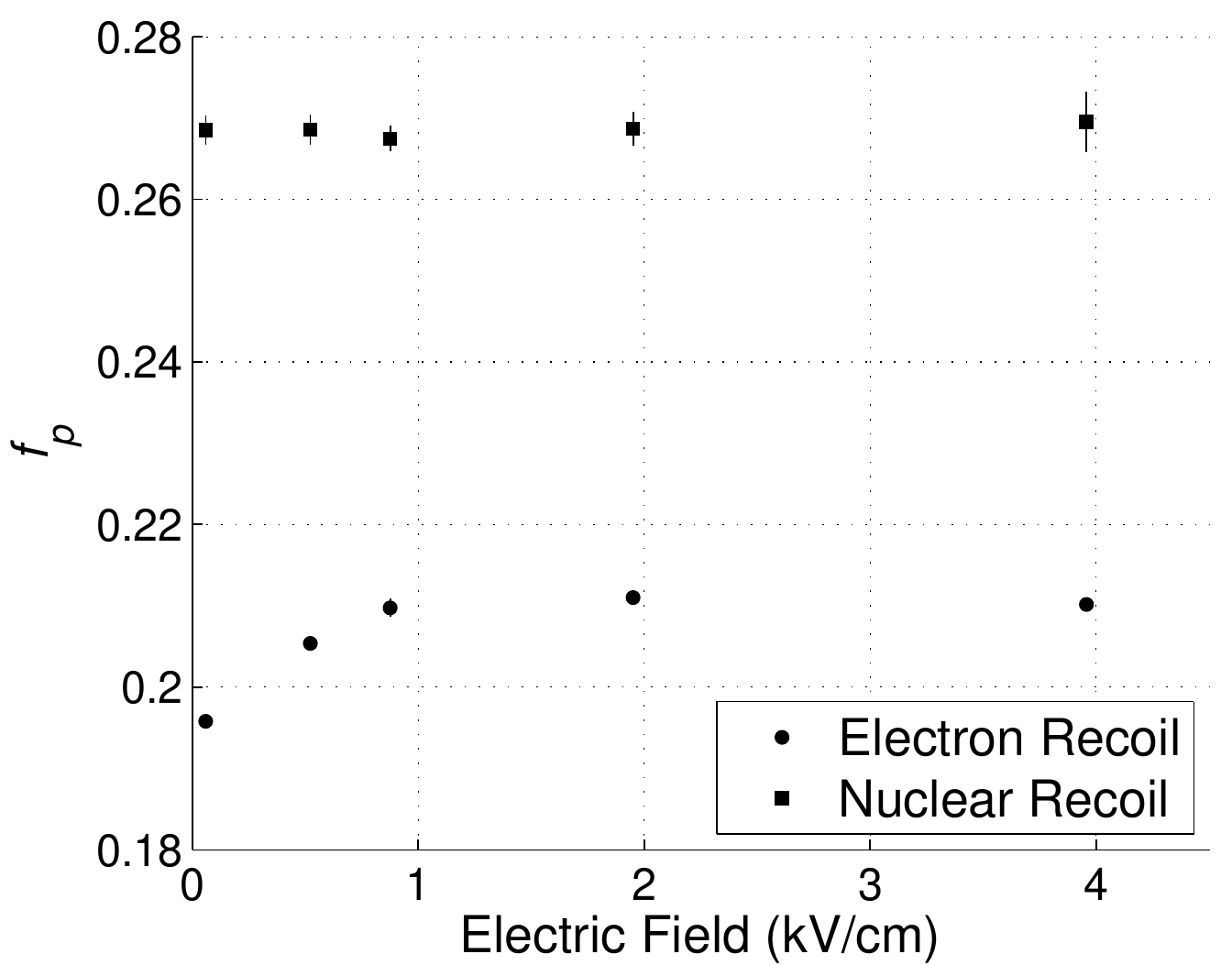}
	\caption{The fit mean value of $\it{f_p}$ between 60 and 80 keVr versus electric field for electron and nuclear recoils.  The error bars represent the statistical uncertainty.  The error bars for the electron recoils are too small to be seen.}
	\label{fig:Fit_Mean_Vs_Field_prompt_total_60_to_80_keVr}
\end{figure}

\begin{figure}[t]
	\centering
	\includegraphics[width=8.0cm]{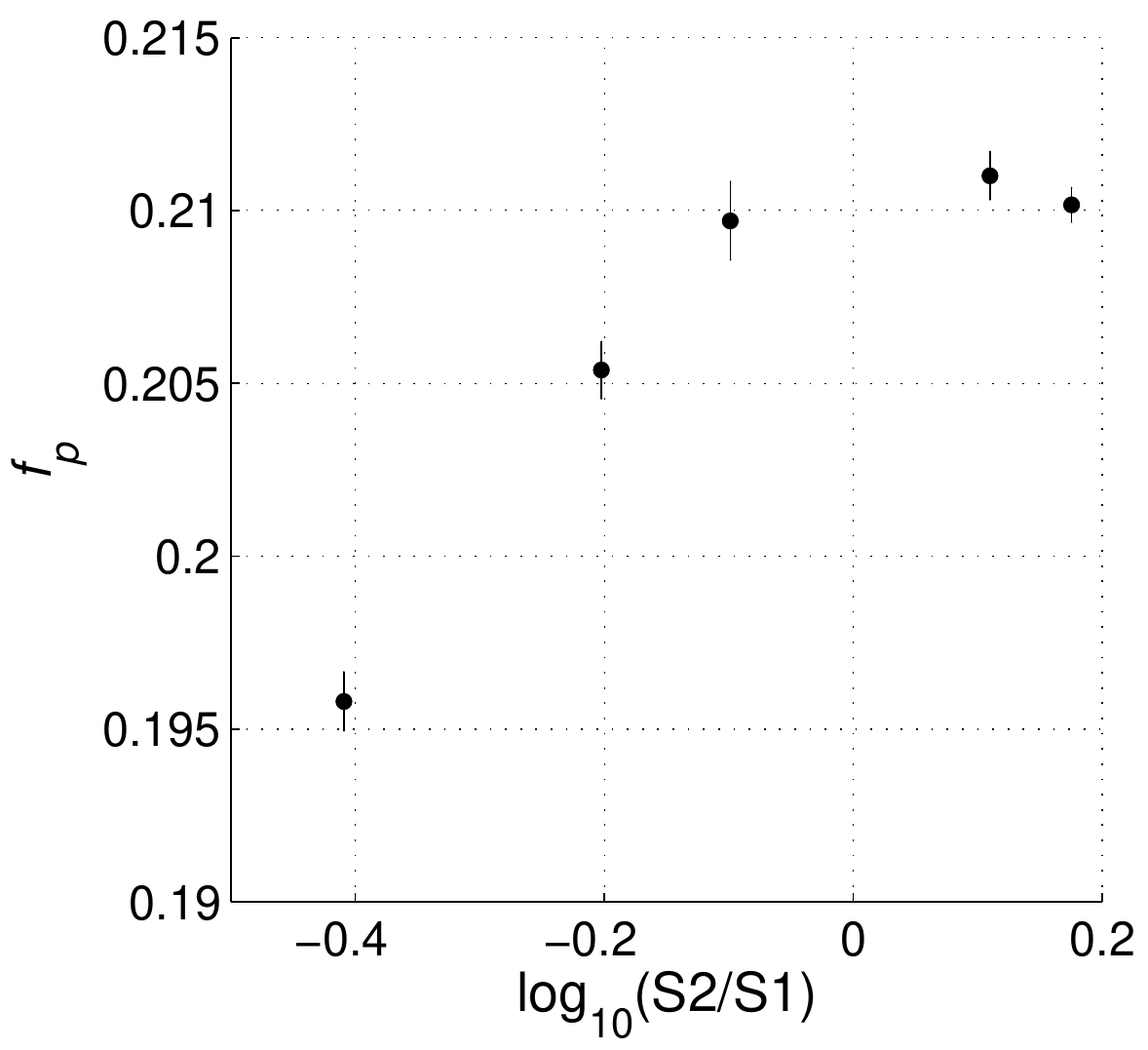}
	\caption{Mean $\it{f_p}$ versus the mean $\log_{10}(S2/S1)$ for electron recoils between 60 and 80 keVr at five different electric fields, with both $f_p$ and $\log_{10}(S2/S1)$ generally increasing with drift field.  The error bars represent the statistical uncertainties.  The error bars for $\log_{10}(S2/S1)$ are too small to be seen.}
	\label{fig:prompt_total_versus_log10_S2_S1_60_to_80_keVr}
\end{figure}

\section{Results}

	Figure \ref{fig:Xed_Leakage_Versus_keVr_0100_V_cm} shows the electron recoil leakage as a function of energy for the data set taken with a drift electric field of 0.06 kV/cm.  This data set demonstrated the best PSD performance among the five taken.  Note that an electron recoil leakage of 0.5 means that there is no discrimination as the nuclear recoil acceptance is 0.5.  Not shown are the leakage values calculated by counting events.  These values agree within error with the Gaussian-extrapolated leakage values, indicating that any anomalous interactions that would have otherwise affected the discrimination are either weak or nonexistent.  (In comparison, the Gaussian-extrapolated and counting leakages for $\log_{10}(S2/S1)$ generally fail to match due to various anomalous events such as those that suffer from charge loss.)  
	The error bars of the leakage include only the statistical fit uncertainties.  The leakage in the prototype is very high below 20 keVr but is $\leq$0.05 for events $\geq$55~keVr.  For comparison, the mean electron recoil leakage with $\log_{10}(S2/S1)$ is $\sim$0.005 below 20~keVr and $\sim$0.002 between 50 and 80~keVr for the window with 50\% nuclear recoil acceptance.  The mean $\it{f_p}$ of the electron recoil events in the $^{133}$Ba and $^{252}$Cf data sets agree within error at 70~keVr.  Systematic shifts in the $\it{f_p}$ of the electron recoils can be as large as the statistical uncertainties.
	
	Figure \ref{fig:band_separation_sigma_versus_energy_0100_V_cm} shows the separation in $\it{f_p}$ and the actual and Monte Carlo electron recoil band widths as a function of energy.  These plots reveal that the rise in leakage with decreasing energy is due to both a convergence of the two bands and an increase in the band width.
	The merging of the electron and nuclear recoil bands can be explained by the convergence in $dE/dx$ and drop in recombination light for electron recoils.  (The electronic stopping powers of alpha, electron and nuclear recoils are given in \cite{Shutt:2007zz}.)  The small differences between the actual and Monte Carlo widths indicate that any intrinsic contribution to the fluctuations in $\it{f_p}$ must be relativity small and that statistical fluctuations dominate.

	Figure \ref{fig:Leakage_Vs_Field_prompt_total_60_to_80_keVr} displays the field dependence of the leakage between 60 and 80 keVr.  The leakage increases as the electric field is increased.  The band width changes negligibly with the electric field since the intrinsic component is very small and the statistical contribution does not change as the number of photoelectrons per unit recoil energy remains roughly the same.  The trend in the leakage is due to the change in the mean $\it{f_p}$ of the electron recoils as seen in Figure \ref{fig:Fit_Mean_Vs_Field_prompt_total_60_to_80_keVr}.  The mean $\it{f_p}$ of the electron recoils possesses a strong field dependence below $\sim$1 kV/cm which is similar to the behavior of the electron recoil charge yield \cite{Aprile:2006kx}.  Figure \ref{fig:prompt_total_versus_log10_S2_S1_60_to_80_keVr} shows the positive correlation between mean $\it{f_p}$ and $\log_{10}(S2/S1)$ (a measure of the charge yield) between 60 and 80 keVr.  In contrast, the charge yield and pulse shape of nuclear recoils both change little with field \cite{Aprile:2006kx}.  This strongly suggests that charge recombination plays a central role in the pulse shape.  Although, a positive correlation is observed for the mean values across different electric fields, the correlation for events at an electric field is roughly zero.  The correlation is either nonexistent or is masked by the statistical fluctuations.

\begin{figure}[t]
\centering
\includegraphics[width=8.6cm]{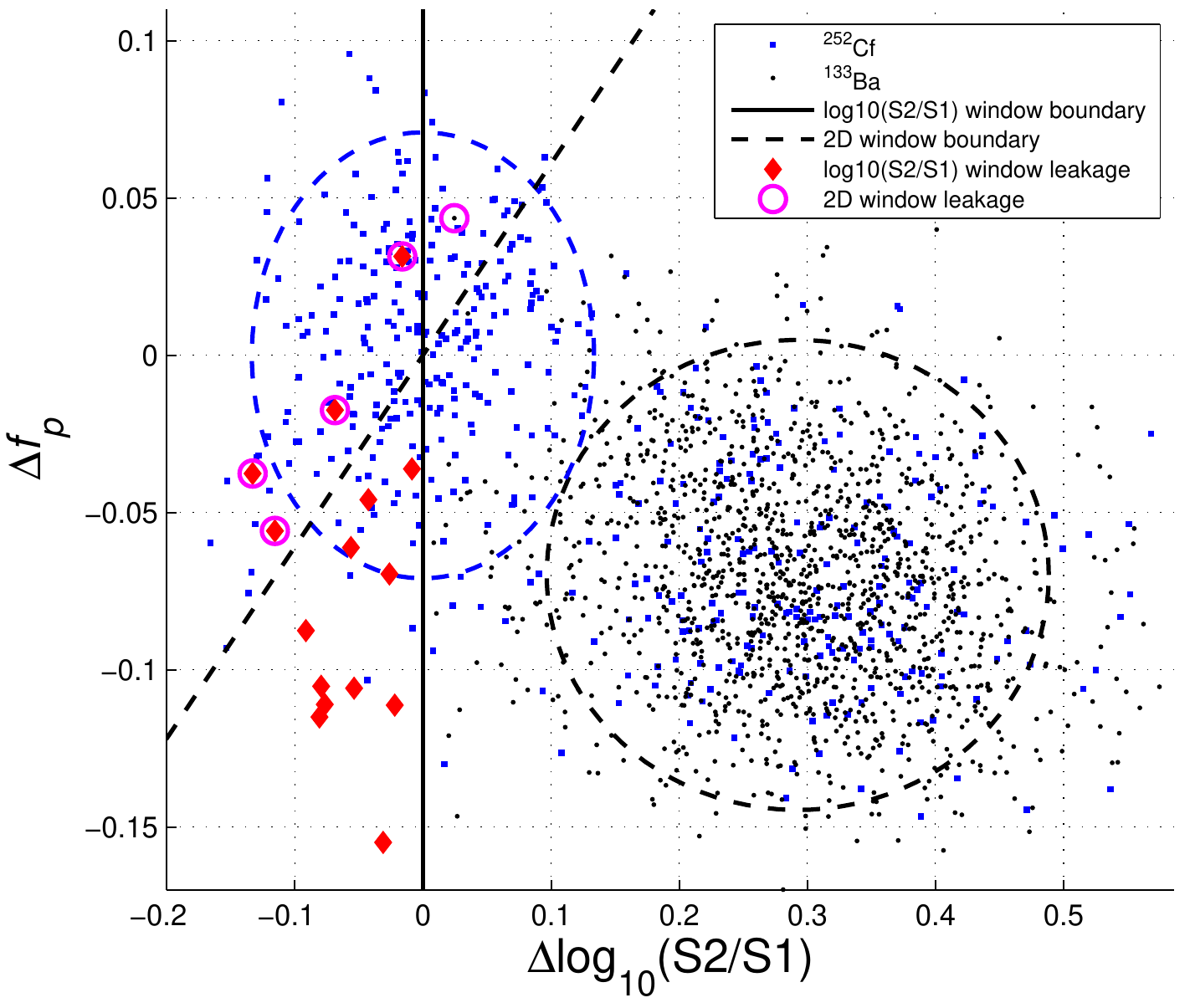}
\caption{Scatter plots of $\it{f_p}$ versus $\log_{10}(S2/S1)$ (both with respect to the nuclear recoil mean) between 70 and 80 keVr at 0.06 kV/cm with the solid and dashed straight line representing the bounds of the $\log_{10}(S2/S1)$ and 2D windows, respectively.  The red diamonds and magenta circles highlight the leakage events of the $\log_{10}(S2/S1)$ and 2D windows, respectively.  The dashed circles represent the $2\sigma$ ellipses.}
	\label{fig:Prompt_Total_Vs_log10_S2_S1_y_Diag_Window_70_to_80_keVr_0100_V_cm}
\end{figure}

\begin{figure}[t]
	\centering
	\includegraphics[width=8.0cm]{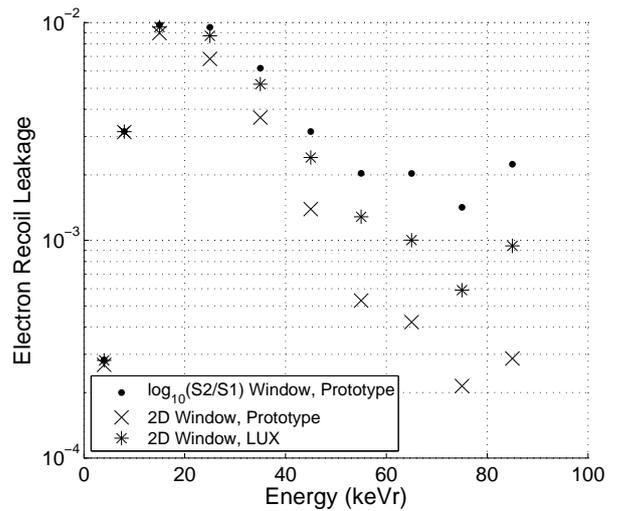}
	\caption{Gaussian-extrapolated electron recoil leakage versus recoil energy at 0.06 kV/cm with the $\log_{10}(S2/S1)$ and 2D windows.}
	\label{fig:2D_Window_Leakage_Vs_keVr_0100_V_cm}
\end{figure}

The electron recoil leakage can be reduced without decreasing nuclear recoil acceptance by changing from the usual nuclear recoil acceptance window defined by $\log_{10}(S2/S1)$ to one defined in $\log_{10}(S2/S1)$-$\it{f_p}$ space (``2D window'').  For simplicity, the lower bound of the 50\% acceptance window is defined as a diagonal straight line through the center of the nuclear recoil cluster.  The optimal angle of the boundary and corresponding leakage are calculated based on the band separations and electron recoil widths in $\log_{10}(S2/S1)$ and $\it{f_p}$ with the assumption that the two quantities are uncorrelated.  Figure \ref{fig:Prompt_Total_Vs_log10_S2_S1_y_Diag_Window_70_to_80_keVr_0100_V_cm} is a scatter plot of $\it{f_p}$ versus $\log_{10}(S2/S1)$ from the $^{133}$Ba and $^{252}$Cf data between 70 and 80 keVr at 0.06 kV/cm with the leakage events of the two windows highlighted.  Figure \ref{fig:2D_Window_Leakage_Vs_keVr_0100_V_cm} shows the Gaussian-extrapolated leakages with the two windows in the prototype (as well as the active volume-averaged projected performance for the larger detector discussed in the next section).  The reduction in the Gaussian leakage is $\sim$6\% below 20 keVr and $\sim$84\% between 60 and 90 keVr.

\begin{figure}[tp]
	\centering
	\includegraphics[width=8.0cm]{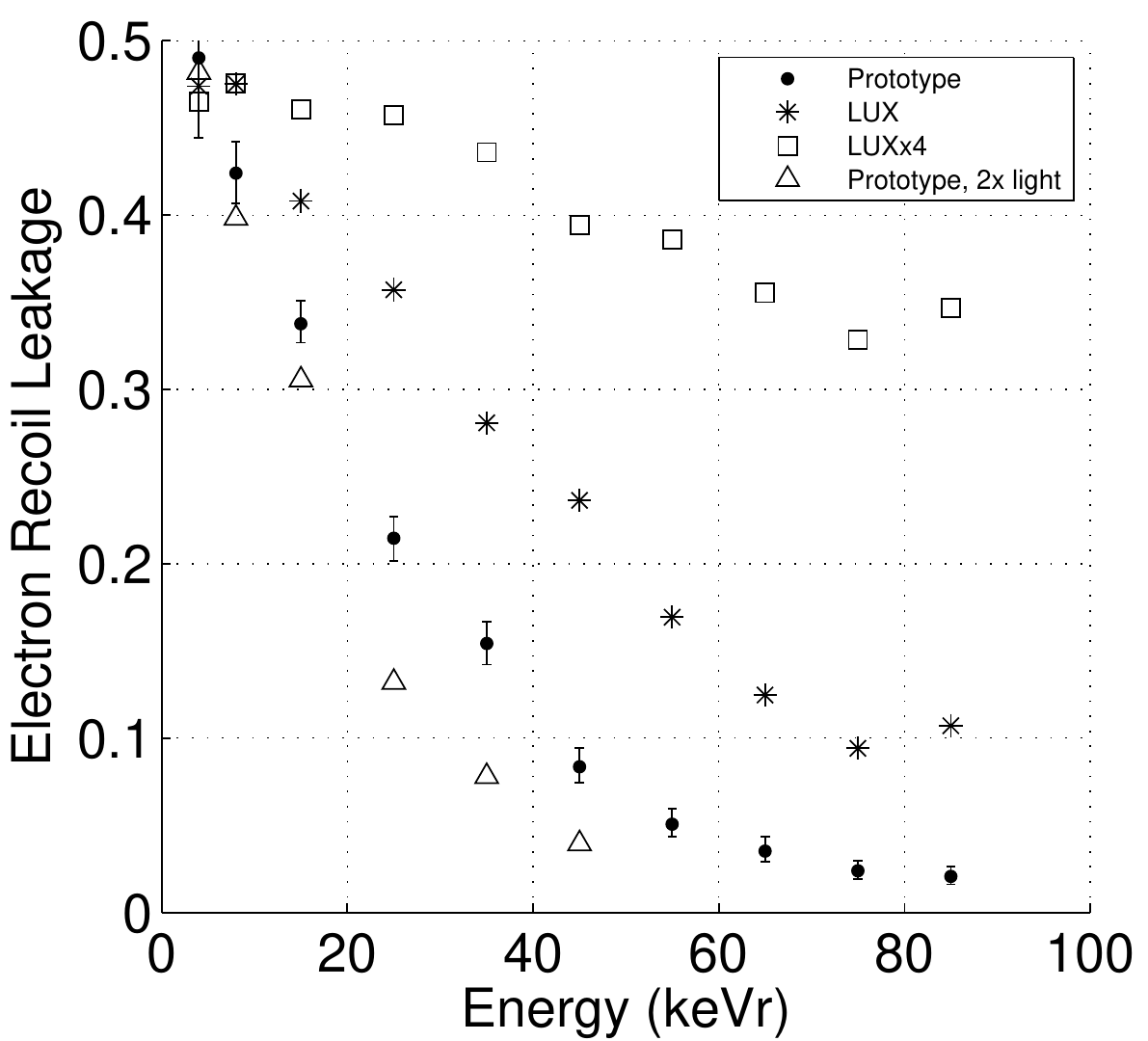}
	\caption{Gaussian leakage as a function of recoil energy for the prototype, the prototype with two times the primary light collection, LUX, and LUXx4 at a drift field of 0.06 kV/cm using the discriminant $\it{f_p}$.  }
	\label{fig:Xed_LUX_LUXx4_Leakage_Versus_keVr_0100_V_cm}
\end{figure}

\section{Pulse Shape Discrimination Projections}

	The results can be used to estimate the discrimination performance of a detector with higher light collection.  The band separation would be the same.  The band width at energy $E$ of a detector with a factor $x$ times the primary light collection of the prototype is the sum (quadrature) of the Monte Carlo width at $x\times E$ plus the intrinsic width (difference between the actual and Monte Carlo widths) at $E$.  Figure \ref{fig:Xed_LUX_LUXx4_Leakage_Versus_keVr_0100_V_cm} shows the projected leakage at 0.06 kV/cm for the prototype with two times the current light collection efficiency.	Although increasing the light collection efficiency significantly decreases the band width, the leakage below 20 keVr is still rather high because of the small and immutable band separation.

	In addition to simulating the photon propagation for the prototype, a next generation two-phase xenon detector called the Large Underground Xenon Detector, or LUX, and a detector with four times the dimensions of LUX (LUXx4) were also simulated.  Nominal LUX active volume dimensions of 49 cm diameter and 54 cm height were used.  Both detectors are presumed to have the same light collection as the prototype.  The time-of-flight distributions for fifty $(r,z)$ coordinates were obtained.  At each electric field, 24,000 S1 pulses of each recoil type were simulated between 0 and 100 keVr at random points in the detectors.  For these simulated pulses, $t_{window} = 36$ ns is used in the definition of $\it{f_p}$.  Figure \ref{fig:Xed_LUX_LUXx4_Leakage_Versus_keVr_0100_V_cm} shows the detector-averaged electron recoil leakage in LUX, LUXx4, and the prototype.  The projections for LUX and LUXx4 use the widths as predicted by the Monte Carlo.  The PSD performance is very weak for LUXx4 at all energies and thus PSD would probably never be used in such a large detector except for events near the PMTs whose photons, to a large degree, strike the PMTs directly.
  
  Also projected were the leakages in LUX and LUXx4 using the 2D window (Figure
\ref{fig:2D_Window_Leakage_Vs_keVr_0100_V_cm}).  The projections use the fits to $\log_{10}(S2/S1)$ from the real data in conjunction with the fits to $\it{f_p}$ from the Monte Carlo data.  The reduction in leakage for LUX by changing to the 2D window is $\sim$2\% below 20 keVr and $\sim$55\% between 60 and 90 keVr.  The improvement for LUXx4 (not shown) is negligible at all energies.
	
	The use of PSD would clearly not be productive in a large detector searching for elastic WIMP recoils as the efficiency of PSD is weakest at low energies where most of the events are expected to appear.  However, PSD might be useful in inelastic recoil searches.  Models for inelastic interactions predict a spectrum with few or no events at low energies and a peak at $\sim$40~keVr \cite{TuckerSmith:2001hy,TuckerSmith:2004jv,Chang:2008gd}.

\section{Conclusion}

	We have studied the energy and electric field dependence of PSD in a two-phase xenon time projection chamber and have shown that a reduction in leakage could be obtained by using a nuclear recoil acceptance window that utilizes both the pulse shape and charge-to-light discriminants.  A pulse shape Monte Carlo shows that the fluctuations in the pulse shape quantity are primarily statistical in nature and that PSD in larger detectors is severely weakened by fluctuations in the photon time-of-flight.  We also find that improvements in light collection are unlikely to meaningfully improve the discrimination efficiency at lower energies due to the small and unchangeable separation between the electron and nuclear recoil bands.  Although the performance is weak at the energies relevant to searches for elastic WIMP recoils, PSD may be useful in looking for higher energy inelastic WIMP recoils.

\section{Acknowledgments}

The authors would like to thank our XENON collaborators for their assistance.  This work was funded by NSF grants NSF-0502690 and PHY-0400596.

\bibliography{PulseShapePaper_arxiv_Ver1_0}

\end{document}